\documentclass[12pt]{article}
\usepackage{amscd}
\usepackage{verbatim}
\usepackage{amssymb}
\newcommand{\qed}{\hfill $\Box$ \medskip}

\renewcommand{\thesection}{\Roman{section}}

\renewcommand{\theequation}{\thesection.\arabic{equation}}
\begin{document}
\vskip 0 true cm \flushbottom
\begin{center}
\vspace{24pt} { \large \bf A Gradient Flow for Worldsheet
Nonlinear Sigma Models} \\
\vspace{30pt}
{\bf T Oliynyk}$^{\dag}$ \footnote{todd.oliynyk@aie.mpg.de},
{\bf V Suneeta}$^{\sharp\ddag}$ \footnote{suneeta@math.unb.ca},
{\bf E Woolgar}$^{\flat\ \ddag}$
\footnote{ewoolgar@math.ualberta.ca} 

\vspace{24pt} 
{\footnotesize $^\dag$ Max-Planck-Institut f\"ur
Gravitationsphysik (Albert Einstein Institute), Am M\"uhlenberg 1,
D-14476 Potsdam, Germany.\\
$^\sharp$ Dept of Mathematics and Statistics, University of New
Brunswick, Fredericton, NB, Canada E3B 5A3.\\
$^\ddag$ Theoretical Physics Institute, University of Alberta,\\
Edmonton, AB, Canada T6G 2J1.\\
$^\flat$ Dept of Mathematical and Statistical Sciences, University of Alberta,\\
Edmonton, AB, Canada T6G 2G1. }
\end{center}
\date{\today}
\bigskip

\begin{center}
{\bf Abstract}
\end{center}
\noindent We discuss certain recent mathematical advances, mainly
due to Perelman, in the theory of Ricci flows and their relevance
for renormalization group (RG) flows. We consider nonlinear sigma
models with closed target manifolds supporting a Riemannian
metric, dilaton, and 2-form $B$-field. By generalizing recent
mathematical results to incorporate the $B$-field and by
decoupling the dilaton, we are able to describe the 1-loop
$\beta$-functions of the metric and $B$-field as the components of
the gradient of a potential functional on the space of coupling
constants. We emphasize a special choice of diffeomorphism gauge
generated by the lowest eigenfunction of a certain Schr\"odinger
operator whose potential and kinetic terms evolve along the flow.
With this choice, the potential functional is the corresponding
lowest eigenvalue, and gives the order $\alpha'$ correction to the
Weyl anomaly at fixed points of $(g(t),B(t))$. The lowest
eigenvalue is monotonic along the flow, and since it reproduces
the Weyl anomaly at fixed points, it accords with the $c$-theorem
for flows that remain always in the first-order regime. We compute
the Hessian of the lowest eigenvalue functional and use it to
discuss the linear stability of points where the 1-loop
$\beta$-functions vanish, such as flat tori and K3 manifolds.

\setcounter{equation}{0}
\newpage

\section{Introduction}
\setcounter{equation}{0}

\noindent A standard approach to renormalization group (RG) flow
for a quantum field theory consists of deriving differential
equations governing the behaviour of the coupling constants under
changes in the renormalization scale (\cite{Wilson}, \cite{WK},
\cite{WH}). These RG flow equations are written in terms of
$\beta$-functions which are components of a vector field tangent
to the flow on the space of coupling constants of the theory. The
$\beta$-functions can be computed in a loop expansion. For the
worldsheet (i.e., 2-dimensional) bosonic sigma model, the loop
expansion parameter is $\alpha'$, the square of the string scale.

A basic result is Zamolodchikov's $c$-theorem \cite{Zamolodchikov,
Cardy}, which implies that certain RG flows are irreversible. This
theorem asserts the existence of a function on the space of coupling
constants of certain 2-dimensional quantum field theories called the
$C$-function, which decreases monotonically along any
renormalization group trajectory from an unstable to a stable fixed
point, and equals the Weyl anomaly (the central charge of the
Virasoro algebra) at fixed points. The $c$-theorem was extended to
nonlinear sigma models with compact target spaces by Tseytlin
\cite{Tseytlin}. The $C$-function obeys the monotonicity formula
$dC/dt = -\kappa(\beta,\beta)$ for $t$ a parameter along the flow,
$\kappa$ a non-negative quadratic form on the space of coupling
constants, and $\beta$ the array of $\beta$-functions of the
coupling constants. The $c$-theorem is not contingent on the loop
expansion for $\beta$.

A longstanding question is whether RG flow is a gradient flow: Is
the vector field defined by the $\beta$-functions orthogonal to
level surfaces of a potential function on the space of coupling
constants. Recent advances in mathematics have shed light on this
matter. Consider the special case of a nonlinear sigma model whose
target space is purely gravitational (in particular, the
anti-symmetric $B$-field is absent), and with $\beta$ replaced by
its 1-loop (order $\alpha'$) approximation. Then the 1-loop RG flow
for the target space metric is known in the mathematics literature
as Ricci flow \cite{Hamilton}. This flow was long known to be
gradient on a space of coupling constant endowed with a metric of
indefinite sign \cite{Friedan}. But the above discussion of the
$c$-theorem and $\kappa$ suggests there may be a positive-definite
metric for 1-loop gradient flow, from which one could obtain a
monotonicity formula. Perelman \cite{Perelman} has now shown that
this flow on closed (target) manifolds of arbitrary dimension is in
fact a gradient flow on a space of coupling constants with
positive-definite metric.

Then what do these recent mathematical advances mean for RG flows
of sigma models more generally, when $B$ is not held to zero? This
paper is intended to address this issue. We restrict our attention
to the first of many potentially relevant results announced in
\cite{Perelman}, the gradient nature of the flow for $g$ and
related monotonicity. We endeavour to set out this result in some
detail in language appropriate to the RG setting. Our first task
is therefore to generalize it to incorporate not only the target
manifold's Riemannian metric $g$ but also the anti-symmetric
2-form field $B$ and dilaton $\Phi$ (though the dilaton can often
be ignored by decoupling it from the system using a suitably
chosen diffeomorphism). We find that for sigma models whose target
space is a closed manifold, the order $\alpha'$ RG flow of $(g,B)$
is gradient on a space of coupling constants with
positive-definite metric, and thus the flow is
irreversible.\footnote
{Elsewhere, we have extended Perelman's work to noncompact
asymptotically flat manifolds with somewhere-negative scalar
curvature and zero $B$-field \cite{OSW}.}
The irreversibility argument is easiest when we choose a certain
diffeomorphism gauge along the flow, and then we call the
potential function $\lambda$. We note that the full RG flow (i.e.,
not the order $\alpha'$ truncation) on closed target manifolds is
known to be irreversible---this is a consequence of the
$c$-theorem. However, in the absence of knowledge of the higher
loop corrections to the $\beta$-function, it is not practically
possible to compute the $C$-function on closed target manifolds.
Our result can be considered valid when the 1-loop truncation of
the $\beta$-function is a good approximation. In this case, we can
explicitly compute the potential function $\lambda$ that generates
the 1-loop RG flow without requiring knowledge about regimes where
``stringy'' (higher order in $\alpha'$) corrections become
important.

We find the value of the potential function $\lambda$ at fixed
points to be non-negative, and zero at any Ricci-flat fixed point.
This raises the undesirable possibility that a Ricci-flat fixed
point such as a flat torus might flow to a non-Ricci-flat one. To
preclude this possibility, we compute the Hessian of the potential
and use the resulting second variations to discuss the linear
stability, as well as the rigidity (or isolation), of Ricci-flat
fixed points. We confirm linear stability for the particular
examples of flat tori and K3 manifolds. Our considerations lead us
to briefly discuss the possible rigidity of Ricci-flat
perturbative string vacua, at which the $\beta$-functions vanish
to all orders in $\alpha'$. This may be of interest when viewed in
the light of investigations into the topology of the configuration
space of string theory \cite{Vafa, DMW}.

In Section II, we discuss the sigma model under consideration,
recalling its RG flow equations at order $\alpha'$ and decoupling
the dilaton from the flow of $g$ and $B$.

Section III is the heart of the paper, particularly the first
subsection, wherein we describe and generalize Perelman's
approach. Perelman's remarkable insight (see also \cite{FOZ, FO})
is that the Ricci flow of $g$, modified by the correct choice of
diffeomorphism, becomes the gradient flow of a potential $\lambda$
which is actually the lowest eigenvalue of a certain Schr\"odinger
operator on the target manifold,\footnote
{In fact, the relevance of this eigenvalue problem for RG flow was
proposed in \cite{FOZ}, in the special case of zero $B$-field and
a closed, 2-dimensional target manifold. It was noticed in
\cite{FO} that the work of \cite{Perelman} validates this
proposal.}
and we show that this carries over {\it mutatis mutandis} to the
$(g,B)$ flow as well. This leads to an easy proof of the absence
of periodic or homoclinic behaviour (i.e., the
``irreversibility'') of the 1-loop flow. In the second subsection,
we evaluate this eigenvalue at a fixed point. We show that it is
$\ge 0$ and is $-4$ times the order $\alpha'$ correction to the
Weyl anomaly at fixed points. The final subsection contains a
brief aside on the gradient nature of RG flow with an arbitrary
diffeomorphism; i.e., not a diffeomorphism chosen as above to
connect to the Schr\"odinger problem.

Section IV contains a derivation of the second variation of the
potential function at a fixed point. This section builds on a
similar result for Ricci flow presented in \cite{CHI}. In Section
V we apply the second variation formula to discuss linear
stability of fixed points, including some particular Ricci-flat
examples. We conclude Section V with some speculative remarks
concerning the potential applicability of our results to an issue
in string theory. Section VI contains remarks on higher-order
flows and the $C$-theorem.

An Appendix contains some calculations related to Section III that
we believe would unnecessarily clutter the main text. That said,
we have tried to provide a certain level of calculational detail,
especially when such detail has not been provided in the
mathematics literature.

Throughout, the target manifold is closed (i.e., compact and without
boundary) and is assumed to carry a positive-definite Riemannian
metric. Our RG flow equations agree with those that appear in
\cite{Polchinski}.

\section{RG Flow of the Worldsheet Nonlinear Sigma Model}
\setcounter{equation}{0}

\noindent The 2-dimensional, or worldsheet, nonlinear sigma model
is a quantum field theory of maps $X^i$ from a 2-dimensional
Riemannian manifold $(\Sigma,h)$, the {\it worldsheet}, to another
Riemannian manifold $(M,g)$, the {\it target manifold} or {\it
target space}, which herein we take to be a closed manifold of dimension
at least $3$. We let $\epsilon$ denote the volume element on the
worldsheet and let $R(h)$ be the worldsheet scalar curvature. The
sigma model action is (using coordinates
$\sigma^{\alpha}=(\sigma^1,\sigma^2)$ on $\Sigma$)
\begin{equation}
S=-\frac{1}{\alpha'}\int_{\Sigma} d^2\sigma \left [ \sqrt{h}
h^{\alpha\beta} g_{ij} \partial_{\alpha} X^i \partial_{\beta} X^j
+ \epsilon^{\alpha\beta}B_{ij}\partial_{\alpha} X^i
\partial_{\beta} X^j -\alpha'\sqrt{h}\Phi R(h) \right ]
\label{eq2.1} \ ,
\end{equation}
where $g_{ij}$, $B_{ij}=-B_{ji}$, and $\Phi$ are the target space
metric, $B$-field, and dilaton, respectively. This model describes
the motion of a bosonic string in a background wherein the
massless string modes have acquired vacuum expectation values
$g_{ij}$, $B_{ij}$, and $\Phi$. The action is invariant under
reparametrizations and conformal rescalings on the worldsheet and
under the addition $B\mapsto B+d\omega$ of an exact form $d\omega$
to $B$, which is a target space 2-form. The gauge-invariant 3-form
field strength for $B$ is $H:=dB$.

Cut-off independence of the regulated quantum theory leads to
renormalization group flow equations (see, e.g.,
\cite{Polchinski})
\begin{eqnarray}
\frac{\partial g_{ij}}{\partial t} &=& -\alpha' \left ( R_{ij} +
2\nabla_i \nabla_j \Phi -\frac{1}{4}H_{ikl}H_j{}^{kl}\right ) \ ,
\label{eq2.2}\\
\frac{\partial H}{\partial t} &=& \alpha' \left ( \frac{1}{2}
\Delta_{\rm LB} H -d \langle H,{\rm grad} \Phi \rangle \right ) \
, \label{eq2.3}\\
\frac{\partial \Phi}{\partial t} &=& -A+\alpha' \left (
\frac{1}{2} \Delta \Phi - \vert \nabla \Phi \vert^2 +\frac{1}{24}
\vert H \vert^2 \right ) \ , \label{eq2.4}
\end{eqnarray}
where $\Delta$ is the Laplacian, $\Delta_{\rm LB} H := -(d\delta +
\delta d) H$ is the Laplace-Beltrami operator acting on the 3-form
$H:=dB$, $A$ is a constant whose value depends on the target
manifold dimension (and the number of ghost fields when they are
present), and $t$ is the log of the renormalization
scale.\footnote
{We do not presume $t$ to be in any way related to physical time
here, although it has been conjectured that RG flow could model
real time evolution in certain situations.}

Equation (\ref{eq2.3}) is actually derived by taking the curl of
the flow equation
\begin{eqnarray}
\frac{\partial B_{ij}}{\partial t} &=& \alpha' \left (
\frac{1}{2}\nabla^k H_{kij} -H_{kij}\nabla^k \Phi
+\nabla_i\omega_j-\nabla_j\omega_i\right ) \nonumber\\
&=&-\alpha' \left ( \frac{1}{2}\delta H +\langle H,{\rm grad} \Phi
\rangle +d\omega\right )_{ij} \ . \label{eq2.5}
\end{eqnarray}
The $\omega$ terms arise because $B$ is determined only up to the
exterior derivative of an arbitrary 1-form $\theta$. Then
$\frac{\partial B}{\partial t}$ acquires a contribution which,
because exterior differentiation commutes with
$\frac{\partial}{\partial t}$, can be written as the exterior
derivative of the 1-form $\omega:=\frac{\partial \theta}{\partial
t}$.

These equations are written with respect to a coordinate basis
fixed with respect to $t$. In a basis that changes with $t$, extra
terms will be introduced into the evolution equations. We will
exploit this now to decouple $\Phi$, and later to demonstrate the
monotonicity formula.

Pulling back by the $t$-dependent diffeomorphism $\varphi_t$
generated by the vector field $\alpha' {\rm grad}\Phi$ adds a Lie
derivative term to each of the flow equations. For example, the
left-hand side of the equation for $g_{ij}$ becomes $\varphi_t^*
\frac{\partial}{\partial t}g_{ij}=(\frac{\partial}{\partial
t}{\tilde g}_{ij}-\pounds g_{ij})\circ\varphi_t
=(\frac{\partial}{\partial t}{\tilde
g}_{ij}-2\alpha'\nabla_i\nabla_j\Phi)\circ\varphi_t$, where
${\tilde g}_{ij}:=\varphi_t^* g_{ij}$. The right-hand side is
natural under diffeomorphisms and becomes, schematically,
$\varphi_t^* ({\rm RHS})={\rm RHS}\circ\varphi_t$. The other
evolution equations transform similarly, yielding
\begin{eqnarray}
\frac{\partial {\tilde g}_{ij}}{\partial t} &=& -\alpha' \left (
{\tilde R}_{ij} -\frac{1}{4}{\tilde H}_{ikl}{\tilde
H}_j{}^{kl}\right ) \ ,
\label{eq2.6}\\
\frac{\partial {\tilde H}}{\partial t} &=& \frac{\alpha'}{2}
\Delta_{\rm LB}{\tilde H}\ ,
\label{eq2.7}\\
\frac{\partial {\tilde \Phi}}{\partial t} &=& -A+\alpha' \left (
\frac{1}{2} \Delta {\tilde \Phi} +\frac{1}{24} \vert {\tilde H}
\vert^2 \right ) \ . \label{eq2.8}
\end{eqnarray}
We say these equations are expressed in {\it Hamilton gauge} in
recognition of the relationship of (\ref{eq2.6}) to the Ricci flow
equation of R Hamilton \cite{Hamilton}. Notice that ${\tilde
\Phi}$ has now decoupled from the evolution equations for ${\tilde
g}_{ij}$ and $\tilde H_{ijk}$.

It will prove convenient to express these equations in an {\it
arbitrary} $t$-dependent coordinate system; {\it i.e.}, to pull
back by an arbitrary $t$-dependent diffeomorphism. To do so, we
simply add Lie derivative terms to each equation. We are
interested in particular in diffeomorphisms generated by the
gradient of a scalar, so for later convenience we choose to write
the generator as $-\alpha'\nabla \psi$ where $\psi$ is arbitrary.
Then the evolution equations become
\begin{eqnarray}
\frac{\partial g_{ij}}{\partial t} &=& -\alpha' \left ( R_{ij}
+\nabla_i \nabla_j \psi -\frac{1}{4}H_{ikl}H_j{}^{kl}\right )
=:-\beta^g_{ij} \ ,
\label{eq2.9}\\
\frac{\partial H_{ijk}}{\partial t} &=& \frac{\alpha'}{2} \left (
\Delta_{\rm LB} H -d\langle H,{\rm grad}\psi\rangle \right )_{ijk}
=:-\beta^H_{ijk}\ ,
\label{eq2.10}\\
\frac{\partial \Phi}{\partial t} &=& -A+\frac{\alpha'}{2} \left (
\Delta \Phi -\nabla\Phi\cdot\nabla \psi +\frac{1}{12} \vert H
\vert^2 \right )=:-\beta^{\Phi} \ . \label{eq2.11}
\end{eqnarray}
The right-hand sides of these equations define what are called
$\beta$-functions. We have dropped the tildes now, but note that
there is of course a distinction between quantities such as
$g_{ij}$ appearing in equations (\ref{eq2.2}--\ref{eq2.4}) and
those in (\ref{eq2.9}--\ref{eq2.11}). Namely, the former are
obtained from the latter by choosing the gauge $\psi=2\Phi$.
Hamilton gauge (denoted by the tildes) is the choice $\psi=0$.

In the arbitrary gauge, (\ref{eq2.5}) becomes
\begin{eqnarray}
\frac{\partial B_{ij}}{\partial t} &=& \frac{\alpha'}{2} \left [
\nabla^k H_{kij} -H_{kij}\nabla^k \psi\right ]\nonumber\\
&& \qquad + 2\alpha' \nabla_{[i} \left [
\omega_{j]}-B_{j]k}\nabla^k \left ( \Phi-\frac{1}{2}\psi\right )
\right ] \ , \label{eq2.12}
\end{eqnarray}
with square brackets on indices indicating anti-symmetrization.
However, we will fix the evolution of the ``internal gauge'' by
imposing the flow equation
\begin{equation}
\frac{\partial B_{ij}}{\partial t} = \frac{\alpha'}{2} \left [
\nabla^k H_{kij} -H_{kij}\nabla^k \psi\right ] =:-\beta^B_{ij} \ .
\label{eq2.13}
\end{equation}
This forces
\begin{equation}
\sigma_i:=\omega_{j}-B_{jk}\nabla^k \left (
\Phi-\frac{1}{2}\psi\right ) \label{eq2.14}
\end{equation}
to be a closed 1-form. We will now show that the flow given by
(\ref{eq2.9}, \ref{eq2.13}) is a gradient flow.

\section{The Gradient Flow}
\setcounter{equation}{0}

\subsection{The Monotonicity Formula}

\noindent In this section we elucidate the first two sections of
\cite{Perelman} (see also the detailed notes \cite{KL}) and
generalize that work to sigma models with $B$-field.

The space of coupling constants ${\cal G} \ni
(g_{ij}(x),B_{ij}(x),\Phi(x))$, $x\in M$, factors as ${\cal G}
=G\times C^{\infty}(M)$, where $(g_{ij}(x),B_{ij}(x))\in G$. The
$C^{\infty}$ factor\footnote
{According to the way we have defined points of ${\cal G}$,
strictly speaking it is not $C^{\infty}(M)$ that splits off from
${\cal G}$ but rather a trivial bundle whose sections belong to
$C^{\infty}(M)$.}
will be used to accommodate both the dilaton $\Phi$ and the
diffeomorphism generating function $\psi$, but it is important not
to equate these.\footnote
{That would be a gauge choice; e.g., the choice $\psi=2\Phi$
produces Hamilton gauge (\ref{eq2.2}--\ref{eq2.4}).}

Consider now a section $(g_{ij},B_{ij},\psi)$ of ${\cal G}$, where
we take $g$, $B$, and $\psi$ to be related by (\ref{eq2.9}) and
(\ref{eq2.13}) respectively, but the $t$-evolution of $\psi$ is
arbitrary. If a choice of $t$-evolution for $\psi$ is made,
equations (\ref{eq2.9}) and (\ref{eq2.13}) will then determine an
evolving section $(g(t),B(t))$ in $G$. Because the dilaton $\Phi$
is decoupled, we do not need to compute its $t$-evolution
simultaneously. Rather, we can compute $\Phi(t)$ {\it a
posteriori} from (\ref{eq2.11}), once the $t$-evolutions for
$g(t)$ and $B(t)$ are determined. In this way, each choice of
evolution $\psi(t)$ gives a distinct $t$-evolving parametrization
of the coupling constants $(g_{ij}(t,x),B_{ij}(t,x),\Phi(t,x))$.

In Subsection III.3 we describe, for each choice $\psi$ of
parametrization of the coupling constants, a potential that
generates a gradient flow for $(g(t),B(t))$ on the space
$G\subseteq {\cal G}$ endowed with a natural choice of inner
product. However, by choosing $\psi$ in a certain very natural way
(which we dub {\it Perelman gauge}), the resulting gradient flow
is particularly useful, and it is that choice which we concern
ourselves with first. Define the functional
\begin{equation}
F[g,B,\psi]:=\int_M dV e^{-\psi} \left [ R + \vert \nabla \psi
\vert^2 -\frac{1}{12}\vert H \vert^2 \right ] \ . \label{eq3.1}
\end{equation}
Integrating by parts, we can write
\begin{equation}
F[g,B,\psi]=\int_M dV e^{-\psi/2} \left [ R -\frac{1}{12}\vert H
\vert^2 -4\Delta \right ] e^{-\psi/2} \ . \label{eq3.2}
\end{equation}
Now let $u(t,x)$, $x\in M$, be the lowest eigenfunction of the
Schr\"odinger operator $R -\frac{1}{12}\vert H \vert^2 -4\Delta$.
The operator depends on $t$ through the flowing metric $g(t)$, and
thus so do the eigenfunctions. Normalize $u$ (and the other
eigenfunctions) to unity:
\begin{equation}
\int_M u^2 dV=1\ . \label{eq3.3}
\end{equation}
Since the lowest eigenfunction $u$ has no nodes, it has a
well-defined logarithm. We use this to define a function $P$ by
\begin{equation}
e^{-P(t,x)/2}:=u(t,x)\ .\label{eq3.4}
\end{equation}
Then, for $\lambda$ the eigenvalue belonging to $u$, we have
\begin{eqnarray}
&&\left [ R -\frac{1}{12}\vert H \vert^2 -4\Delta \right ] u =:
\lambda u\ \nonumber\\
&\Rightarrow& R-\frac{1}{12}\vert H \vert^2 +2\Delta P - \vert
\nabla P \vert^2 = \lambda \ . \label{eq3.5}
\end{eqnarray}
Clearly, the choice $P=\psi$ minimizes the functional
(\ref{eq3.2}) over all $C^{\infty}(M)$ functions obeying $\int_M
e^{-\psi} dV =1$. Thus on $G$ there is a new functional
$\lambda[g,B]$, equal in value at $(g(t),B(t))$ to $\lambda(t)$,
defined by
\begin{eqnarray}
\lambda[g,B]&:=&\inf_{\{\psi|\int_M e^{-\psi} dV=1\} }
F[g,H,\psi]\ , \label{eq3.6} \\
\Rightarrow \lambda(t) &=& \lambda[g(t),B(t)]\ .\nonumber
\end{eqnarray}
and the infimum is realized by the choice $\psi=P$.

We can compute the gradient of $\lambda[g,B]$ on $G$ by evaluating
the first variation of $F[g,B,\psi]$, requiring $\psi$ to be a
solution $\psi=P$ of (\ref{eq3.5}) all along the variation. The
first step is to compute the free variation in which $\psi$ is not
constrained. This variation is familiar in physics. If we
momentarily think of $\psi$ as the dilaton\footnote
{keeping in mind of course that it is $\Phi$, not $\psi$, that
obeys the dilaton RG flow; indeed, when $\psi$ is constrained in
the manner of Subsection III.3, it can be shown to evolve in $t$
according to a {\it backwards} parabolic evolution equation.}
for purposes of computing this variation, then $F$ is the low
energy effective action in string theory and its variational
derivative is well-known. For completeness, we provide a
derivation in the Appendix; cf.\ equation (\ref{eqA.9}). The
variation gives
\begin{eqnarray}
\frac{dF}{ds}&=&\int_M \biggl [ \left ( -R^{ij}-\nabla^i \nabla^j
\psi +\frac{1}{4} H^i{}_{kl} H^{jkl} \right ) \frac{\partial
g_{ij}}{\partial s}\nonumber \\
&&\qquad + \left ( R -\frac{1}{12}\vert H \vert^2 + 2\Delta \psi
-\vert \nabla \psi \vert^2 \right ) \left ( \frac{1}{2} g^{ij}
\frac{\partial g_{ij}}{\partial s} - \frac{\partial \psi}{\partial
s} \right ) \nonumber \\
&&\qquad + \frac{1}{2} \left ( \nabla_k H^{kij} - H^{kij} \nabla_k
\psi \right ) \frac{\partial B_{ij}}{\partial s} \biggr ]
e^{-\psi} dV \ . \label{eq3.7}
\end{eqnarray}

Now impose the constraint that, for each value of $s$ along the
variation, $\psi$ is not freely varied but rather is fixed to obey
(\ref{eq3.4}). That is, $\psi(s)=P(s)= -2\log u(s)$, where $u(s)$
is the lowest eigenfunction of the Schr\"odinger operator
determined by the {\it varied} metric and $B$-field. Then
(\ref{eq3.7}) becomes the first variation formula for
$\lambda[g,B]$. As well, on the right-hand side of (\ref{eq3.7})
we use first (\ref{eq3.5}) and then (\ref{eq3.3}) and
(\ref{eq3.4}) to write
\begin{eqnarray}
&&\int_M \left ( R -\frac{1}{12}\vert H \vert^2 + 2\Delta P -\vert
\nabla P \vert^2 \right ) \left ( \frac{1}{2} g^{ij}
\frac{\partial g_{ij}}{\partial s} - \frac{\partial P}{\partial s}
\right ) e^{-P}dV\nonumber \\
&=&\lambda \int_M \left ( \frac{1}{2} g^{ij} \frac{\partial
g_{ij}}{\partial s} - \frac{\partial P}{\partial s}
\right ) e^{-P}dV\nonumber \\
&=&\lambda \frac{d}{ds} \int_M e^{-P}dV\nonumber\\
&=& 0\ . \label{eq3.8}
\end{eqnarray}
Thus, the middle line of (\ref{eq3.7}) vanishes and we are left
with
\begin{eqnarray}
\frac{d\lambda}{ds}&=&\int_M \biggl [ \left ( -R^{ij}-\nabla^i
\nabla^j P +\frac{1}{4} H^i{}_{kl} H^{jkl} \right ) \frac{\partial
g_{ij}}{\partial s}\nonumber \\
&&\qquad + \frac{1}{2} \left ( \nabla_k H^{kij} - H^{kij} \nabla_k
P \right ) \frac{\partial B_{ij}}{\partial s} \biggr ] e^{-P} dV \
. \label{eq3.9}
\end{eqnarray}

Now we can consider the elements of $G$ to be 2-index tensors with
symmetric part $g$ and skew part $B$. Then the inner product is
given by
\begin{eqnarray}
\left \langle T,T' \right \rangle &:=& \int_M e^{-P} g^{ik}g^{jl}
T_{ij} T'_{kl} dV\nonumber\\
&=&\int_M e^{-P} g^{ik}g^{jl}\left [ S_{ij}S'_{kl} + A_{ij}
A'_{kl} \right ] dV \ , \label{eq3.10}
\end{eqnarray}
where $S$ is the symmetric part of $T\in TG$ and $A$ is the
anti-symmetric part. Then (\ref{eq3.9}) is the directional
derivative $\langle T, {\rm Grad\ } \lambda \rangle$ of $\lambda$
in the direction $T:=\left ( \frac{\partial g}{\partial s},
\frac{\partial B}{\partial s}\right )$. Then we can read off the
gradient.

\bigskip

\noindent{\bf Proposition 3.1.} {\sl The RG flow for $(g,B)$, with
diffeomorphism gauge $\psi$ fixed to be a solution $P$ of
(\ref{eq3.5}), is the gradient flow in $G$ generated by the
potential $\lambda[g,B]$:
\begin{eqnarray}
\frac{\partial}{\partial t} \left (
\begin{array}{c}
g_{ij}\\
B_{ij}
\end{array}
\right ) &\equiv& \left (
\begin{array}{c}
-\alpha' \left ( R_{ij}
+\nabla_i \nabla_j P -\frac{1}{4}H_{ikl}H_j{}^{kl}\right )\\
\frac{\alpha'}{2} \left [ \nabla^k H_{kij} -H_{kij}\nabla^k P
\right ]
\end{array}
\right )\nonumber\\
&=& \alpha'{\rm \ Grad\ } \lambda[g,B] \ , \label{eq3.11}
\end{eqnarray}
and $\lambda(t)$ is monotone increasing along the gradient flow:
\begin{eqnarray}
\frac{d\lambda}{dt}&=&\alpha'\int_M \biggl [ \left \vert
R_{ij}+\nabla_i \nabla_j P -\frac{1}{4} H_{ikl} H_j{}^{kl} \right
\vert^2\nonumber \\
&&\qquad + \frac{1}{4} \left \vert \nabla^k H_{kij} - H_{kij}
\nabla^k P \right \vert^2 \biggr ] e^{-P} dV \ . \label{eq3.12}
\end{eqnarray}
Furthermore, fixed points of (\ref{eq3.11}) (where $H=dB$) are
stationary points of $\lambda$.}

\bigskip

\noindent{\bf Proof.} The gradient formula (\ref{eq3.11}) can be
read off from (\ref{eq3.9}). To obtain (\ref{eq3.12}), consider
the special case in (\ref{eq3.9}) of a variation in $(g,B)$
produced by evolving $(g,B)$ along the RG flow. That is, let
$\left ( \frac{\partial g}{\partial s}, \frac{\partial B}{\partial
s}\right )$ in (\ref{eq3.9}) be given by the flow equations
(\ref{eq2.9}, \ref{eq2.13}) with $s=t$. Finally, for fixed points
of (\ref{eq3.11}), the right-hand side of (\ref{eq3.12}) vanishes.
\qed

\bigskip

\noindent{\bf Corollary 3.2.} {\sl $\lambda$ is monotonic along RG
flow (\ref{eq2.9}, \ref{eq2.13}).}

\bigskip

\noindent{\bf Proof.} Starting from the same initial data, the
resulting solutions of the gradient flow (\ref{eq3.11}) and the RG
flow with arbitrary $\psi$ (\ref{eq2.9}, \ref{eq2.13}) are related
by at worst a time-dependent diffeomorphism. But $\lambda(t)$ is
diffeomorphism-invariant, and monotonic along the gradient
flow.\qed

\bigskip

We now discuss briefly the absence of periodic 1-loop RG flows. We
will call a solution of the flow (\ref{eq2.9}, \ref{eq2.13}) a
{\it breather} if it is periodic up to gauge and diffeomorphism;
i.e., if there is a diffeomorphism $\varphi$, a 1-form $\omega$,
and parameter values $t_1<t_2$ such that
$(g(t_1),B(t_1))=(\varphi^*g(t_2), \varphi^*B(t_2)+d\omega)$. A
solution that is not a breather is a {\it homoclinic orbit} if it
is eternal (i.e., defined for all $t\in(-\infty,\infty)$) with
$(g(t),B(t))$ converging to $(g_0,B_0)$ for $t\to -\infty$ and to
$(\varphi^*g_0,\varphi^*B_0+d\omega)$ for $t\to +\infty$.

\bigskip
\noindent {\bf Proposition 3.3.} {\sl There are no periodic or
homoclinic orbits of the RG flow other than the fixed points of
(\ref{eq3.11}).}

\bigskip

\noindent {\bf Proof.} For a periodic orbit of the flow
(\ref{eq2.9}, \ref{eq2.13}), there will be some $t_1<t_2$ such
that $\lambda(t_1)=\lambda(t_2)=:\Lambda$. Then by monotonicity,
$\lambda(t)=\Lambda$ for all $t\in[t_1,t_2]$. But by
(\ref{eq3.12}), this can only happen if the right-hand side
of(\ref{eq3.11}) vanishes throughout $[t_1,t_2]$. This is the
condition for a fixed point (with $\psi=P$). For the homoclinic
case, note that the sequence $\lambda(nT)-\lambda(-nT)$, $n\in
\mathbb Z$, $T>0$, is increasing. However, for a homoclinic orbit,
this sequence must converge to zero. Therefore
$\lambda(nT)-\lambda(-nT)=0$ for all $n$. Setting $t_1=-nT$,
$t_2=nT$, we see as before that the flow on $[t_1,t_2]=[-nT,nT]$
is at a fixed point. But we can take $n$ arbitrarily large.\qed

\bigskip

Note that this result implies that if the flow equations with any
potential $\psi$ have a fixed point in which the right-hand sides
of (\ref{eq2.9}) and (\ref{eq2.13}) vanish, {\it then the
diffeomorphism generator $\psi$ must be a solution $P$ of}
(\ref{eq3.5}).

Perelman, in Section 2 of \cite{Perelman}, was able to go farther.
Using $\lambda$ as we have defined it, but of course with $B=0$,
he was able to prove the non-existence of nontrivial expanding
breathers, metrics that are equal at two different $t$-values up
to a diffeomorphism and a homothety such that the metric with
greater $t$-value has greater volume. However, the argument does
not go through generally if the $B$-field is permitted to have
nonzero field strength $H$ at some time during the flow.

\subsection{$\lambda$ and the Weyl Anomaly at Fixed Points}

\noindent We can now evaluate the eigenvalue $\lambda$ at a
stationary point, thus at a fixed point of the flow for which $H$
is an exact 3-form. Since we consider a one-parameter family of
flows with parameter $s$, we write $\lambda_s$ for the eigenvalue
and denote the stationary point by $s=0$. At such a point, each
term in (\ref{eq3.12}) vanishes, and thus in particular we must
have
\begin{equation}
R+\Delta P - \frac{1}{4}|H|^2=0\ . \label{eq3.13}
\end{equation}
When this holds, (\ref{eq3.5}) takes the form
\begin{equation}
\lambda_0=\Delta P - \vert \nabla P \vert^2 + \frac{1}{6} \vert H
\vert^2 \ . \label{eq3.14}
\end{equation}
Multiply this by $e^{-P}$ and integrate. On the left-hand side,
$\int_M \lambda_0 e^{-P} dV = \lambda_0 \int_M e^{-P} dV =
\lambda_0$, and on the right-hand side the derivatives of $P$
vanish upon integration by parts (since $M$ is closed). This
yields
\begin{equation}
\lambda_0=\frac{1}{6}\int_M |H|^2 e^{-P} dV\ . \label{eq3.15}
\end{equation}
Notice now that, from the right-hand side of (\ref{eq2.11}) with
the diffeomorphism now chosen so that $\psi=P$ as must be the case
at a fixed point, we can write
\begin{equation}
\int_M e^{-P}\beta^{\Phi} dV = A-\frac{\alpha'}{24} \int_M e^{-P}
\vert H \vert^2 dV = A - \frac{\alpha'}{4}\lambda_0 \ .
\label{eq3.16}
\end{equation}
(Recall that $A$ is a constant depending on the dimension of the
target manifold and the number of ghost fields, if any are
present.) We compare this expression to Tseytlin \cite{Tseytlin}
by considering the combination\footnote
{Tseytlin's definition of $t$ differs from ours by a sign, but the
signs in his RG equations are such that his $\beta$-functions
agree with ours.}
\begin{equation}
{\tilde \beta}:=\beta^{\Phi}-\frac{1}{4}g^{ij} \beta^g_{ij} \ ,
\label{eq3.17}
\end{equation}
which equals the Weyl anomaly at fixed points of g and B
\cite{CFMP}, where of course $\beta^g_{ij}$ vanishes and thus we
obtain
\begin{equation}
\int_M e^{-P}{\tilde \beta} dV = A-\frac{\alpha'}{4} \lambda_0\ .
\label{eq3.18}
\end{equation}
But at fixed points, $\tilde\beta$ is constant on $M$
(\cite{CFMP}; for the case where no $B$-field is present, this
follows as an integrability condition for the fixed point equation
$\beta^g_{ij}=0$ as discussed in \cite{CP}; for the case with
$B$-field, see the discussion in \cite{Osborn}). Then
(\ref{eq3.18}) reduces to the relation
\begin{equation}
\tilde\beta=A-\frac{\alpha'}{4} \lambda_0 \ \label{eq3.19}.
\end{equation}

An example is provided by a fixed point of $(g(t),B(t))$ with
$|H|^2:=H_{ijk}H^{ijk}$ constant over the closed manifold $M$ and
with dilaton linear in the scale $t$. We must first note that
(\ref{eq3.14}) can now be written as
\begin{equation}
\Delta e^{-P} = \frac{1}{6}e^{-P} \left ( \vert H \vert^2 - \int_M
\vert H \vert^2 e^{-P} dV \right ) \ . \label{eq3.20}
\end{equation}
Since $|H|^2$ is constant over $M$ (and so by (\ref{eq3.15})
$|H|^2=6\lambda$), the right-hand side of (\ref{eq3.20}) vanishes
and $e^{-P}$ is harmonic. Then $P$ is constant. (For the $B=0$
case, the result was already known from work of Bourguignon
\cite{Bourguignon}.) Then (\ref{eq2.11}) reduces to
\begin{equation}
\frac{\partial \Phi}{\partial t} = -A+\frac{\alpha'}{2} \left (
\Delta \Phi + \frac{1}{2} \lambda \right ) = -\beta^{\Phi} \
,\label{eq3.21}
\end{equation}
A solution is given by
\begin{eqnarray}
\Phi&=&\left ( \frac{\alpha'}{4} -A\right ) t+\Phi_0\ , \label{eq3.22}\\
\frac{\partial \Phi_0}{\partial t} &=& \frac{\alpha'}{2}\Delta
\Phi_0\ , \label{3.23}
\end{eqnarray}
where the operator $\Delta$ has no time dependence, in consequence
of our being at a fixed point with trivial diffeomorphism $\nabla
P =0$. Then if we choose that $\Phi_0$ is constant in time, it
also will be harmonic and thus spatially constant at all times,
and we obtain, as claimed, that
\begin{equation}
{\tilde \beta} = \beta^{\Phi}= A - \frac{\alpha'}{4}\lambda\ .
\label{eq3.24}
\end{equation}

\subsection{A Gradient Flow with Arbitrary Diffeomorphism Term}

\noindent The RG flow (\ref{eq2.9}, \ref{eq2.13}) with {\it
arbitrary} diffeomorphism not necessarily determined by
(\ref{eq3.5}) is nonetheless a gradient flow as well. To see this,
return to (\ref{eq3.7}) but do not require now that $\psi=P=-2\log
u$ at each $s$. Instead, to eliminate the middle line, choose any
fiducial measure $dm:= e^{-\psi}dV$ and hold it fixed pointwise
along the variation. That is,

\bigskip

\noindent{\bf Proposition 3.4.} {\sl The RG flow (\ref{eq2.9},
\ref{eq2.13}) is the gradient flow of $F[g,B,\psi]$ along the
surface in ${\cal G}$ determined by the fixed fiducial measure
$dm:=e^{-\psi} dV$ on $M$.}

\bigskip

\noindent{\bf Proof.} If $dm$ is fixed then $g^{ij}\frac{\partial
g_{ij}}{\partial s} -\frac{\partial \psi}{\partial s}=0$ and the
middle line on the right-hand side of (\ref{eq3.7}) again
vanishes. We obtain (\ref{eq3.9}), from which the gradient can be
read off. Since $dm$ can be chosen arbitrarily, $\psi$ is now
arbitrary as well.\qed

\bigskip

While the approach of Proposition 3.4 may seem more general and
perhaps simpler than the approach of Proposition 3.1, it is in
fact far less powerful, because $F$ depends on the arbitrary
diffeomorphism potential $\psi$. To prove results such as
Proposition 3.3, one must remove the $\psi$ dependence by passing
to $\lambda$ which, in contrast to $F$, is geometrically
meaningful---and has a clear physical interpretation.

\section{The Second Variation of $\lambda$}
\setcounter{equation}{0}

\noindent In the case of Ricci flow ($B=0$) the second variation
formula for $\lambda_s$ was written down in \cite{CHI}. We
generalize the formula here for arbitrary $B$-field. The next
section applies this formula in a special case.

In an endeavour to minimize clutter, we establish a convention. If
a quantity below is to be considered as a function of $s$, we
write $s$ explicitly as an argument or subscript. If $s$ does not
appear, the quantity is evaluated at $s=0$ (possibly after $s$
differentiation; this should be clear from context). We will use
$\nabla_k$ to denote the covariant derivative compatible with
$g_{ij}(s)$ and $D_k$ for the covariant derivative compatible with
$g_{ij}\equiv g_{ij}(0)$.

The second variation formula about an arbitrary point is
complicated. Fortunately, for most purposes, we only need the
second variation about a stationary point of $\lambda_s$. Thus we
require the $s=0$ fields $(g,H)$ to obey
\begin{eqnarray}
R_{ij}+D_i D_j P -\frac{1}{4} H_{ikl} H_j{}^{kl}&=&0\ ,
\label{eq4.1} \\
D^k H_{kij} - H_{kij} D^k P &=&0 \ . \label{eq4.2}
\end{eqnarray}
Thus from (\ref{eq3.9}) the second variation formula about a
stationary point is
\begin{eqnarray}
\frac{d^2\lambda_s}{ds^2}\bigg \vert_{s=0}&=&\int_M \biggl [
h^{ij} \frac{\partial}{\partial s}\left ( -R_{ij}-\nabla_i
\nabla_j
P +\frac{1}{4} H_{ikl} H_j{}^{kl} \right ) \nonumber \\
&&\quad + \frac{1}{2} \beta^{ij}\frac{\partial}{\partial s} \left
( \nabla^k H_{kij} - H_{kij} \nabla^k P \right ) \biggr ] e^{-P}
dV \ , \label{eq4.3}
\end{eqnarray}
where we define
\begin{eqnarray}
h_{kl}&:=& \frac{\partial g_{kl}}{\partial s}\quad , \quad
h^{ij}:=g^{ik}g^{jl}h_{kl} \ , \label{eq4.4}\\
\beta_{kl}&:=&\frac{\partial B_{kl}}{\partial s}\quad , \quad
\beta^{ij} :=g^{ik}g^{jl}\beta_{kl}\ , \label{eq4.5}
\end{eqnarray}
and for use below
\begin{equation}
Q:=\frac{\partial P}{\partial s}\ . \label{4.6}
\end{equation}

The standard formulas
\begin{eqnarray}
\frac{\partial}{\partial s}R_{ij} &=& \nabla_k
\frac{\partial}{\partial s}\Gamma^k_{ij} -\nabla_i
\frac{\partial}{\partial s}\Gamma^k_{jk}\label{eq4.7}\\
\frac{\partial}{\partial s}\Gamma^k_{ij} &=& \frac{1}{2} g^{kl}
\left ( \nabla_i h_{jl} +\nabla_j h_{il}-\nabla_l h_{ij} \right )
\label{eq4.8}
\end{eqnarray}
yield the easy identity (writing $D$ for $\nabla|_{s=0}$ whenever
we can and using a standard result for the variation of a
Christoffel symbol)
\begin{eqnarray}
&&\frac{\partial}{\partial s}\left ( -R_{ij}
-D_iD_j P \right ) \nonumber\\
&=&-e^P D_k \left ( e^{-P} \frac{\partial}{\partial s}
\Gamma^k_{ij} \right ) - \left ( D_i D_j \frac{\partial
P}{\partial s} -D_i\frac{\partial}{\partial s}
\Gamma^k_{jk} \right ) \nonumber\\
&=&-e^P D_k \left ( e^{-P} \frac{\partial}{\partial s}
\Gamma^k_{ij} \right ) + D_iD_j \left ( \frac{1}{2} g^{kl}h_{kl}
-Q \right
) \nonumber\\
&=&-\frac{1}{2}e^P g^{kl}D_k \left [ e^{-P} \left ( D_i h_{jl} +
D_j h_{il} - D_l h_{ij} \right ) \right ] \nonumber \\
&&+D_iD_j \left ( \frac{1}{2} g^{kl}h_{kl} -Q \right )
\nonumber\\
&=&-\frac{1}{2} \biggl ( D_i D^k h_{jk} +D_j D^k
h_{ik} -R_{kilj}h^{kl}-R_{kjli}h^{kl}\nonumber \\
&&\quad +R^k_i h_{jk} +R^k_j h_{ik} -\Delta h_{ij}\biggr ) +
D_iD_j \left ( \frac{1}{2} g^{kl}h_{kl} -Q \right )
\nonumber\\
&=& -\frac{1}{2} \biggl ( D_i D^k h_{jk} +D_j D^k h_{ik}
-\Delta_L h_{ij}\biggr )\nonumber\\
&&\quad + D_iD_j \left ( \frac{1}{2} g^{kl}h_{kl} -Q \right ) \ ,
\label{eq4.9}
\end{eqnarray}
where $\Delta_L$ denotes the Lichnerowicz Laplacian
\begin{equation}
\Delta_L h_{ij}:=\Delta h_{ij}+R_{kilj}h^{kl}+R_{kjli}h^{kl}-R^k_i
h_{jk} -R^k_j h_{ik} \ . \label{eq4.10}
\end{equation}
Thus we obtain (round brackets around indices indicate
symmetrization):

\bigskip

\noindent{\it The Second Variation Formula:}
\begin{eqnarray}
\frac{d^2\lambda_s}{ds^2}\bigg \vert_{s=0}&=& \frac{1}{2} \int_M
e^{-P} h^{ij}\left ( \Delta_L h_{ij} -H_{ikm}
H_{jl}{}^m h^{kl} \right ) dV\nonumber\\
&&+\int_M e^{-P}h^{ij}\left [ -D_{(i}D^kh_{j)k}+D_iD_j \left (
\frac{1}{2}g^{kl}h_{kl} -Q \right ) \right ] dV\nonumber\\
&&+\frac{1}{2}\int_M e^{-P}\beta^{ij}\frac{\partial}{\partial s}
\left ( \nabla^k H_{kij} - H_{kij} \nabla^k P \right ) dV \ ,
\label{eq4.11}
\end{eqnarray}
for variations about a general stationary point.

\bigskip

As a check on our results, we restrict to variations $(h,0)$ in
which $B$ is fixed. We determine $Q$ in the second variation
formula by noting that (\ref{eq3.5}) must hold all along the
variation (i.e., for all $s$). Differentiate it and evaluate the
derivative at $s=0$. On the left-hand side, $\frac{d}{ds}\lambda_s
\vert_{s=0}=0$ since $s=0$ is a stationary point, while the
right-hand side can be simplified by using (\ref{eq4.7}), etc, to
obtain
\begin{equation}
D^i \left [ e^{-P} D_i \left ( g^{ij}h_{ij}-2Q \right ) \right ] =
D^i D^j \left ( e^{-P} h_{ij} \right ) \ . \label{eq4.12}
\end{equation}
We write the solution of this equation (one always exists and is
unique modulo an inconsequential additive constant, since $M$ is
closed) as
\begin{equation}
v_h:= g^{ij}h_{ij}-2Q\ , \label{eq4.13}
\end{equation}
This determines $Q$, given $h_{ij}$. Now we further restrict to
the case where $|H|^2$ is constant on the manifold. Then by
(\ref{eq3.19}), we can set $P=0$. Under these circumstances, the
second variation formula becomes
\begin{eqnarray}
\frac{d^2\lambda_s}{ds^2}\bigg \vert_{s=0}&=& \frac{1}{2} \int_M
h^{ij}\left ( \Delta_L h_{ij} -H_{ikm}
H_{jl}{}^m h^{kl} \right ) dV \ , \label{eq4.14}\\
&&+\int_M \left ( \vert {\rm div\ }(h)\vert^2 -\frac{1}{2}\vert
Dv_h \vert^2 \right ) dV \ , \nonumber
\end{eqnarray}
where $({\rm div\ }h)_i:=D^kh_{ik}$, and this yields agreement
with the Ricci flow result of \cite{CHI} when $H_{ijk}=0$. For
$h_{ij}$ {\it transverse} (i.e., if $D^i h_{ij}=0$), {\it both}
terms in the second integrand vanish.

\section{$H=0$ Fixed Points}
\setcounter{equation}{0}

\noindent Consider a family of solutions of the RG flow with $t$
the parameter along each flow and $s$ the family parameter. In the
sequel, we always choose $s$ such that $s=0$ is a stationary point
of $\lambda_s$ and thus a fixed point of the flow. If the second
variation of $\lambda_s$ about $s=0$ is positive at $t=0$, then
$C:=\lambda_s(0) > \lambda_0(0)$ for some $s>0$. Since
$\lambda_s(t)$ is monotonic in $t$ and $\lambda_0(t)$ is constant,
then $\lambda_s(t)\ge C>\lambda_0(0)$ for all $t$. Since
$\lambda_s$ is continuous on $G$, the coupling constants cannot
approach the fixed point couplings along the flow. Thus,
$\frac{d^2}{ds^2}\lambda_s \vert_{s=0}>0$ indicates an instability
of the fixed point.

Now Ricci-flat fixed points have $\lambda=0$, which is the least
possible value of $\lambda$ at a fixed point. A particularly
worrisome scenario would occur if certain Ricci-flat manifolds,
such as flat tori and K3 manifolds, were unstable against small
perturbations and could flow to other, non-Ricci-flat fixed
points. To preclude this possibility, we must study the
eigenvalues of the Hessian of $\lambda_s$ about $s=0$. This notion
of stability is called {\it linear stability}.

It is difficult to study general variations $(h,\beta)$ in both
$g$ and $B$ about an arbitrary fixed point, owing to the
difficulty in diagonalizing the Hessian. However, Ricci-flat fixed
points necessarily have $H=0$, and for a fixed point with $H=0$
things simplify considerably, making it possible. As well as the
direct simplification of setting $H=0$, the argument surrounding
(\ref{eq3.19}) then gives that $P$ can be set to zero as well, and
so the fixed point condition gives that $R_{ij}=0$. We use
(\ref{eq4.11}) to write
\begin{equation}
\frac{d^2\lambda_s}{ds^2}\bigg \vert_{s=0}=\frac{1}{2} \int_M
\left ( h^{ij}\Delta_L h_{ij} +2 \vert {\rm div\ }(h) \vert^2 -
\vert Dv_h \vert^2 -\frac{1}{3} \vert d\beta \vert^2 \right ) dV\
, \label{eq5.1}
\end{equation}
where now $v_h$ solves
\begin{equation}
D^kD_k v_h \equiv D^kD_k \left ( g^{ij}h_{ij} -2Q \right )=D^i D^j
h_{ij}\ . \label{eq5.2}
\end{equation}

Equation (\ref{eq5.1}) has the form
\begin{eqnarray}
\frac{d^2\lambda_s}{ds^2}\bigg \vert_{s=0}&=&\int_M \left (
\langle h , {\cal L} h \rangle -\frac{1}{6} \vert d\beta
\vert^2 \right ) dV \nonumber\\
&=&\int_M \left ( \langle h , {\cal L} h \rangle -\frac{1}{6}
\langle \beta, d^* d \beta\rangle \right ) dV\, \label{eq5.3}\\
{\cal L} h&:=&\frac{1}{2}\Delta_L h + {\rm div}^*{\rm div\ }h + DD
v_h \ , \label{eq5.4}
\end{eqnarray}
where we have suppressed the indices, written $({\rm div\ } h)_i:=
D^k h_{ik}$, and let ${\rm div}^*$ denote the adjoint of the
divergence and $d^*$ denote the adjoint of $d$ with respect to the
inner product (\ref{eq3.10}). It is known \cite{GIK,Sesum} that
eigenvectors of ${\cal L}$ belonging to positive eigenvalues, if
any exist, are transverse and traceless, and moreover for $h_{ij}$
transverse traceless then
\begin{eqnarray}
{\cal L}\vert_N&=&\frac{1}{2}\Delta_L\vert_N\ ,\label{eq5.5}\\
N&:=&\{h_{ij}\ \vert\ ({\rm div}h)_i:=D^kh_{ik}=0,\ {\rm tr}_g h
:= g^{ij}h_{ij}=0 \}\ . \label{eq5.6}
\end{eqnarray}
Furthermore, on the complement of $N$, ${\cal L}<0$ unless
$h_{ij}$ arises from the action of a diffeomorphism or homothetic
rescaling on $g_{ij}$, in which case ${\cal L}=0$.

The $\vert d\beta \vert^2$ term in (\ref{eq5.3}) is obviously
negative semi-definite, and negative if $\beta$ is not closed.
With that and the above comments concerning ${\cal L}$ in mind, we
define a fixed point $(g,d\beta)$ with $R_{ij}=0$, $H_{ijk}=0$ to
be {\it linearly stable} if $\Delta_L\vert_N\le 0$. If the tangent
directions at $(g,d\beta)$ for which $\Delta_L\vert_N = 0$ can be
exponentiated to give a smooth submanifold ${\cal U}$ of
Ricci-flat fixed points in the space of coupling constants, then
we say that the fixed point $(g,d\beta)$ is {\it integrable}.
%
%
We will call this an {\it integral submanifold of Ricci-flat fixed
points}. Since by construction $\Delta_L<0$ in the complement in
$N$ of $T{\cal U}$ at each point along ${\cal U}$, we call ${\cal
U}$ {\it strongly linearly stable}. If ${\cal U}$ is strongly
linearly stable and if neither the spectrum of $\Delta_L$ nor that
of $\delta d:\Lambda^2(M)\to \Lambda^2(M)$ accumulates at zero, we
will call ${\cal U}$ {\it strictly linearly stable}.

\bigskip

\noindent{\bf Lemma 5.1.} {\sl If a strictly linearly stable
integral submanifold ${\cal U}$ consists entirely of a disjoint
set of one or more Ricci-flat fixed points, then those points are
{\it rigid} (isolated): there are no neighbouring fixed points,
Ricci-flat or not, except those obtained by diffeomorphism and/or
homothety.}

\bigskip

\noindent{\bf Proof:} At $p\in{\cal U}$, consider any submanifold
$S$ whose tangent space at $p$ is contained in $N$. Because
$\lambda$ is zero and stationary at $p$ and $\Delta_L$ is bounded
below zero on $TS$, and $\delta d$ is bounded below zero on
$\Lambda^2(M)$ modulo closed forms, then $\lambda<0$ on some
neighbourhood of $p$ in $S$. But fixed points have $\lambda\ge 0$
by (\ref{eq3.15}). \qed

We have no examples of rigid fixed points. However, the proof
generalizes to the case of submanifolds ${\cal U}$ of nonzero
dimension provided what is meant by ``isolated'' is interpreted to
mean that neighbouring fixed points must also belong to ${\cal U}$
are allowed. This brings us to the cases of greatest interest
here:

\bigskip

\noindent{\it Flat Tori:}

\noindent For flat manifolds, it helps to write (\ref{eq5.3}) as
\begin{equation}
\frac{d^2\lambda_s}{ds^2}\bigg \vert_{s=0}=-\frac{1}{2}\int_M
\left ( 3\vert D_{(i} h_{jk)} \vert^2 + \frac{1}{3} \vert d\beta
\vert^2 \right ) dV\ , \label{eq5.7}
\end{equation}
for $h_{ij}\in N$. This is strictly negative unless $\beta$ is
closed and $D_{(i}h_{jk)} =0$, and then $h_{ij}$ is called a
Killing tensor. For tori, the Killing tensors are always linear
combinations of outer products of translation Killing vectors.
These modes correspond to the relative rescaling of distinct
cycles, holding the torus volume fixed. These relative rescalings
give rise to the moduli space of flat structures on the torus,
which is clearly a strongly linearly stable integral submanifold
of Ricci-flat fixed points in the space of coupling constants.
Moreover, it is evident from the triviality of the eigenvalue
problem in this case that the moduli space is in fact strictly
linearly stable.

\bigskip

\noindent{\it K3 Manifolds:}

\noindent The infinitesimal deformations (meaning in this
situation the $h_{ij}\in N$ such that $\Delta_L h_{ij}=0$) of
K\"ahler Ricci-flat metrics on K3 manifolds are known to actually
correspond to Ricci-flat metrics \cite{Todorov,Sesum}. Once again,
these metrics form a submanifold ${\cal E}$ of Ricci-flat fixed
points. It was shown in \cite{GIK} that $\Delta_L<0$ on the
complement of $T{\cal E}$ in $N$. Thus ${\cal E}$ is a strongly
linearly stable integral submanifold of Ricci-flat fixed points.

\bigskip

Although these results on linear stability are strongly
suggestive, they do not demonstrate {\it dynamical stability}
without further technical argument. In the case of Ricci flow
(i.e., no $B$-field), Sesum \cite{Sesum} has found linear and
dynamical stability to be equivalent when the fixed point
satisfies the integrability condition. Thus flat tori and K3
manifolds are dynamically stable under Ricci flow. We expect
similar results will hold when a $B$-field is present. The issue
is presently under investigation. Nonetheless, the picture that
emerges is one where flat tori and K3s are final and not initial
endpoints of the flow (except of course for the trivial case of a
flow that remains always at the fixed point). If flows that end at
these points begin at unstable fixed points, then monotonicity of
$\lambda$ and the fact that $\lambda=0$ at the final endpoint
would imply that those initial fixed points would have
$\lambda<0$, contradicting (\ref{eq3.15}). The resolution of this
apparent paradox is simply that higher order terms $\alpha'$ are
significant for such flows and cannot be neglected.

Finally, consider the stability of the subset of fixed points of
the first-order flow that remain fixed points of the flow to all
orders in $\alpha'$. That is, they receive no `stringy
corrections' at any order in perturbation theory. These are called
{\it perturbative string vacua}, and are described by conformal
field theories. Questions concerning the topology of the space of
such vacua, the dimension of the moduli space, etc., have been
discussed in the string theory literature primarily in the
language of CFTs and their operator content. Because our results
for general order $\alpha'$ fixed points certainly obviously
descend to perturbative string vacua, we have a complementary
picture in which these questions can be phrased in the language of
target manifold geometry. The clearest case would be that of a
zero-dimensional integral submanifold of Ricci-flat perturbative
string vacua. Then Lemma 5.1 would apply directly.\footnote
{Kahler-Ricci-flat K3 manifolds are perturbative string vacua. A
version of Lemma 5 could be proved for the larger integral
submanifolds that occur there. However, we do not know if these
manifolds are strictly linearly stable, even though they are
strongly linearly stable.}
Checking stability would then be a matter of checking the
eigenvalues of $\Delta_L$. If this could be done and if stability
were confirmed, we could infer that the related CFT should have no
relevant or marginal operators. Now while we presently know of no
specific example of such a zero-dimensional manifold of vacua, we
have seen for flat tori that it need not be difficult to draw
conclusions about relevant operators in the CFT even when the
vacuum belongs to a nontrivial integral submanifold.

In \cite{Vafa}, Vafa addresses the question of whether the
$C$-function can serve as a Morse function for the configuration
space of string theory, the hope being that this would shed light
on the topology of this configuration space. In particular, he
points out that the possible existence of nontrivial fixed points
with no relevant or marginal directions---rigid perturbative
string vacua in the language above---raises the question of
whether the configuration space of string theory is connected.

\section{Concluding Remarks}
\setcounter{equation}{0}

\noindent Throughout, we have limited attention to the most
elementary of the monotone quantities for Ricci flow in
\cite{Perelman} and to order $\alpha'$ $\beta$-functions. To do
more would have made this article unwieldy. It is, however,
interesting to ponder whether Perelman's $W$-entropy, reduced
length, and reduced volume have useful analogues for the flows we
have considered. An analogue of the scale invariant $W$-entropy
would allow one to address and probably rule out the possibility
that there may be solutions of the RG flow which are periodic
except for an overall homothetic rescaling \cite{DM}.

It is also natural to ask whether these techniques might show the
RG flow to be gradient and have a monotonicity formula at higher
order in $\alpha'$. Higher-order RG flow equations have a
significant difficulty, which is that nonlinear combinations of
leading-order spatial derivative terms appear in the flow PDEs.
Then the question may be vacuous, in that these PDEs might not
admit any solutions at all. Physics does not require that they do,
since the exact RG flow, by which we mean that the
$\beta$-functions are not approximated using a truncated loop
expansion, can still exist nonetheless. However, if a gradient
flow is found for, say, the order $\alpha'^2$ RG flow, then this
would be an important step in showing existence of solutions of
the PDEs (as streamlines of the gradient flow). Thus, in the
absence of a separate existence proof for solutions there is no
reason to expect to find a suitable gradient flow, but it is the
very absence of such a proof that makes the question more
interesting.

Thinking beyond the loop expansion, we return to the $C$-theorem.
In the present work, we utilized recent mathematical breakthroughs
to say something about order $\alpha'$ physics. These advances
occurred in the context of Ricci flow and attempts to prove
conjectures concerning 3-manifold topology, but seem to echo the
$C$-theorem (cf Section III.2 and \cite{FOZ}). It is intriguing to
ask what may come from a reverse strategy. For example, might the
$C$-theorem lead to a tower of geometric flows with monotonicity
properties and interesting fixed points, order-by-order in
$\alpha'$?\footnote
{This intriguing idea arose in discussion with Gerhard Huisken.}
Can the Ricci flow with surgery be given a physics interpretation
and might the $C$-theorem have something to say about topology of
manifolds? To realize this potential, it seems very important to
understand more deeply the relationship between the $C$-function
and $\lambda$, or perhaps the other monotonic quantities known for
Ricci flow \cite{Perelman} but which we have not discussed herein.

\section{Acknowledgements}

\noindent This research was partially supported by a Discovery
Grant from the Natural Sciences and Engineering Research Council
of Canada. VS was supported by a fellowship from the Pacific
Institute for the Mathematical Sciences. We thank the organizers
of the Dark Side of Extra Dimensions meeting and the staff of the
Banff International Research Station (BIRS), where this work was
begun. EW thanks the Albert Einstein Institute, the Isaac Newton
Institute, and the organizers of the Global Problems in
Mathematical Relativity (GMR) workshop for hospitality. EW also
thanks MT Anderson, Gerhard Huisken, H-P K\"unzle, and Kostas
Skenderis for discussions, Jim Isenberg for help with K3
manifolds, and Joe Polchinski and especially Hugh Osborn for
advice and references concerning the $C$-theorem for 2-dimensional
sigma models.

\setcounter{equation}{0}

\appendix
\section{Appendix} \label{appendix.sec}
\renewcommand{\theequation}{\thesection.\arabic{equation}}
\setcounter{equation}{0}


\noindent Here we compute the first variation in the functional
$F[g,B,\psi]$ that results from a 1-parameter variation of $g$,
$B$, and $\psi$. We denote the parameter by $s$ and compute the
first variation term-by-term, starting from the formula
\begin{eqnarray}
\frac{dF}{ds}&=&\int_M\left ( \frac{\partial R}{\partial
s}+\frac{\partial}{\partial s} \vert \nabla \psi \vert^2
-\frac{1}{12}\frac{\partial}{\partial s}\vert H\vert ^2\right )
e^{\psi}dV\nonumber\\
&&+\int_M \left ( R+\vert \nabla \psi \vert^2 -\frac{1}{12} \vert
H \vert ^2 \right ) \frac{\partial}{\partial s} \left (
e^{-\psi}dV \right )\ . \label{eqA.1}
\end{eqnarray}

\bigskip

\noindent{\bf Lemma A.1.} {\sl For $M$ compact, then}
\begin{eqnarray}
\int_M\frac{\partial R}{\partial s}e^{-\psi}dV &=&-\int_M\biggl
[R^{ij}+\nabla^i\nabla^j \psi -\nabla^i \psi \nabla^j \psi\nonumber\\
&&\qquad - g^{ij} \left ( \Delta \psi -\vert \nabla \psi \vert^2
\right ) \biggr ] \frac{\partial g_{ij}}{\partial s} e^{-\psi}dV\
, \label{eqA.2}
\end{eqnarray}

\bigskip

\noindent {\bf Proof.} Use the standard formula
\begin{equation}
\frac{\partial R}{\partial s}=-R^{ij}\frac{\partial
g_{ij}}{\partial s} +\nabla^i \left [ \nabla^j\frac{\partial
g_{ij}}{\partial s}-\nabla_i\left ( g^{kl}\frac{\partial
g_{kl}}{\partial s}\right ) \right ]\label{eqA.3}
\end{equation}
and integrate by parts twice.\qed

\bigskip

\noindent{\bf Lemma A.2.} {\sl Assume that (\ref{eq3.1}) holds.
Then}
\begin{eqnarray}
&&\int_M \left ( \frac{\partial}{\partial s} \vert \nabla \psi
\vert^2 \right ) e^{-\psi}dV\nonumber\\
&& = \int_M \biggl [ 2 \left ( \vert \nabla \psi\vert^2 -\Delta
\psi \right ) \frac{\partial \psi}{\partial s} - \frac{\partial
g_{ij}}{\partial s} \nabla^i \psi \nabla^j \psi \biggr ]
e^{-\psi}dV\ . \label{eqA.4}
\end{eqnarray}

\bigskip

\noindent {\bf Proof.}
\begin{equation}
\int_M \left ( \frac{\partial}{\partial s} \vert \nabla \psi
\vert^2 \right ) e^{-\psi}dV = \int_M \left [ 2\nabla^i \psi
\nabla_i \frac{\partial \psi}{\partial s} - \frac{\partial
g_{ij}}{\partial s} \nabla^i \psi \nabla^j \psi \right ]
e^{-\psi}dV \label{eqA.5}
\end{equation}
and integrate by parts. \qed

\bigskip

\noindent{\bf Lemma A.3.}
\begin{eqnarray}
&&-\frac{1}{12}\int_M \left ( \frac{\partial}{\partial s} \vert H
\vert^2 \right ) e^{-\psi}dV\nonumber\\
&& = \int_M \biggl [ \frac{1}{4}H^i{}_{kl}H^{jkl}\frac{\partial
g_{ij}}{\partial s}+\left ( \nabla_k H^{kij} - H^{kij}\nabla_k
\psi \right ) \frac{\partial B^{ij}}{\partial s} \biggr ]e^{-\psi}
dV \ . \label{eqA.6}
\end{eqnarray}

\bigskip

\noindent{\bf Proof.} The first term on the right-hand side is
obvious. The second term follows from the fact that
$\frac{\partial}{\partial s}
\partial_{[i} B_{jk]}=\partial_{[i}\frac{\partial}{\partial s}
B_{jk]}=\nabla_{[i}\frac{\partial}{\partial s} B_{jk]}$. Then
\begin{eqnarray}
&&\int_M\left [ -\frac{1}{6}H^{ijk}\frac{\partial}{\partial
s}H_{ijk} \right ] e^{-\psi}dV\nonumber\\
&=& \int_M\left [
-\frac{1}{2}H_{ijk}g^{ip}g^{jq}g^{kl}\nabla_p\frac{\partial
B_{ql}}{\partial s}\right ] e^{-\psi}dV\nonumber\\
&=&\frac{1}{2}\int_M\left [ \left ( \nabla^k H_{kij} -
H_{kij}\nabla^k \psi \right ) g^{ip}g^{jq}\frac{\partial
B_{pq}}{\partial s}\right ] e^{-\psi}dV\label{eqA.7}
\end{eqnarray}
\qed

\bigskip

\noindent{\bf Lemma A.4.}
\begin{eqnarray}
&&\int_M \left ( R+\vert \nabla \psi \vert^2 -\frac{1}{12} \vert H
\vert ^2 \right ) \frac{\partial}{\partial s} \left ( e^{-\psi}dV
\right )\nonumber \\
&&\qquad=\int_M \left ( R+\vert \nabla \psi \vert^2 -\frac{1}{12}
\vert H \vert ^2 \right )\left ( \frac{1}{2}g^{ij}\frac{\partial
g_{ij}}{\partial s}-\frac{\partial \psi}{\partial s}\right )
e^{-\psi} dV \ . \label{eqA.8}
\end{eqnarray}

\bigskip
\noindent{\bf Proof.} Follows from the formula
$\frac{1}{\sqrt{g}}\frac{\partial \sqrt{g}}{\partial s}=
\frac{1}{2}g^{ij}\frac{\partial g_{ij}}{\partial s}$ for the
derivative of a determinant.\qed

\bigskip
\noindent{\bf Proposition A.5.} {\sl For any arbitrary smooth
1-parameter variation of $g$, $B$, and $\psi$, then}
\begin{eqnarray}
\frac{dF}{ds}&=&\int_M \biggl [ \left ( -R^{ij}-\nabla^i \nabla^j
\psi +\frac{1}{4} H^i{}_{kl} H^{jkl} \right ) \frac{\partial
g_{ij}}{\partial s}\nonumber \\
&&\qquad + \left ( R -\frac{1}{12}\vert H \vert^2 + 2\Delta \psi
-\vert \nabla \psi \vert^2 \right ) \left ( \frac{1}{2} g^{ij}
\frac{\partial g_{ij}}{\partial s} - \frac{\partial \psi}{\partial
s} \right ) \nonumber \\
&&\qquad + \frac{1}{2} \left ( \nabla_k H^{kij} - H^{kij} \nabla_k
\psi \right ) \frac{\partial B_{ij}}{\partial s} \biggr ]
e^{-\psi} dV \ . \label{eqA.9}
\end{eqnarray}

\bigskip
\noindent{\bf Proof:} Follows immediately from Lemmata A.1--4.\qed

\end{document}